\documentclass{article}

\usepackage{arxiv}

\usepackage[utf8]{inputenc} 
\usepackage[T1]{fontenc}    
\usepackage{hyperref}       
\usepackage{cleveref}
\usepackage{url}            
\usepackage{booktabs}       
\usepackage{amsfonts}       
\usepackage{nicefrac}       
\usepackage{microtype}      
\usepackage{lipsum}
\usepackage{graphicx}
\graphicspath{ {./figures/} }

\title{Predicting Drug Solubility Using Different Machine Learning Methods - Linear Regression Model with Extracted Chemical Features vs Graph Convolutional Neural Network}

\author{
John Ho\thanks{These authors contributed equally to this work.}\\
  Harvard University\\
  Cambridge, MA 02138 \\
  \and
  \textbf{Zhao-Heng Yin}\footnotemark[1]\\
  University of California, Berkeley\\
  Berkeley, CA 94720
  \and
  \textbf{Colin Zhang}\\
  Carlmont High School\\
  Belmont, CA 94002 \\
  \and
  \textbf{Nicole Guo}\\
  Lawrence Berkeley National Laboratory\\
  Berkeley, CA 94720 \\
 \and
 \textbf{Yang Ha}\\
  Lawrence Berkeley National Laboratory\\
  Berkeley, CA 94720 \\
}

\begin{document}
\maketitle
\begin{abstract}
Predicting the solubility of given molecules remains crucial in the pharmaceutical industry. In this study, we revisited this extensively studied topic, leveraging the capabilities of contemporary computing resources. We employed two machine learning models: a linear regression model and a graph convolutional neural network (GCNN) model, using various experimental datasets. Both methods yielded reasonable predictions, with the GCNN model exhibiting the highest level of performance. However, the present GCNN model has limited interpretability while the linear regression model allows scientists for a greater in-depth analysis of the underlying factors through feature importance analysis, although more human inputs and evaluations on the overall dataset is required. From the perspective of chemistry, using the linear regression model, we elucidated the impact of individual atom species and functional groups on overall solubility, highlighting the significance of comprehending how chemical structure influences chemical properties in the drug development process. It is learned that introducing oxygen atoms can increase the solubility of organic molecules, while almost all other hetero atoms except oxygen and nitrogen tend to decrease solubility.
\end{abstract}


\section{Introduction}
\label{sec:intro}
In the pharmaceutical industry, the process of discovering new drugs is costly and time-intensive. To reduce expenses and expedite the process, there is usually an early-stage high-throughput screening (HTS) to eliminate molecules that lack desired properties \cite{pereira2007origin}. One such property is solubility, which governs drug uptake, movement, and metabolism in humans \cite{wen2015drug}.

The prediction of molecular solubility, whether based on theoretical principles or experimental data, has been a prominent research field for decades. In 1968, Hansch et al. discovered that the octanol-water partition coefficient (P) can be used for solubility prediction \cite{hansch1968extrathermodynamic}. Subsequently, the Yalkowsky group introduced a general solubility equation (GSE), which incorporated P and the melting point (MP) \cite{yalkowsky1980solubility}. Later, Jorgensen and Duffy utilized Monte Carlo (MC) simulations to predict aqueous solubility by considering structural features such as molecular weight (MW), volume, solvent accessible surface area (SASA), hydrogen bond (HB) counts, and other physical descriptors like water Coulomb and van der Waals interactions (ESXL), as well as hydrophobic and hydrophilic components. Their approach achieved reasonable predictive accuracy on a dataset of 150 organic molecules \cite{jorgensen2000prediction}.

In recent years, with the fast growth of computing power and the development of new algorithms, researchers can now work with more extensive datasets and employ sophisticated machine learning (ML) models \cite{llinas2019solubility}. Several databases, such as AQUASOL and PHYSPROP, used by Huuskonen et al. \cite{huuskonen2000estimation}, ESOL by Delaney \cite{delaney2004esol}, and various solubility handbooks \cite{guillory2003handbook}, have provided access to experimental solubility data for thousands of chemicals. AqSolDB is a newly developed database that combines multiple existing datasets \cite{sorkun2019aqsoldb}. From a methodological perspective, rather than relying on traditional regression models and classic neural network (NN) models, the Barzilay group applied graph convolutional neural networks (GCNN) for molecular property prediction. These GCNNs transform molecular structures into graphs, which can be input into a directed message-passing neural network, achieving state-of-the-art performance \cite{yang2019analyzing}. Moreover, research has extended beyond drug solubility in aqueous solutions to include solute types like small proteins \cite{wirawan2014hector} or various organic solvents \cite{chinta2019machine}.

While these advanced ML algorithms deliver remarkable performance, they often present challenges for human scientists seeking mechanistic insights into the chemistry behind these solubility models. These models are commonly labeled as "black boxes" because it remains difficult to understand the inner workings of, for instance, a 20-layer deep learning NN or a GCNN when all molecules are represented by extensive matrices. From a chemist's perspective, there is a growing need to shift the focus away from performance metrics and toward gaining deeper chemical insights. In this study, our goal is not to solely push the boundaries of predictive accuracy but to harness the strengths of both classical and modern, sophisticated models to enhance our comprehension of the relationship between molecular structures and their chemical properties. With this knowledge, we aim to develop future ML models that combine high accuracy with human interpretability.

\begin{figure}[!htb]
  \centering
  \includegraphics[scale = 0.9]{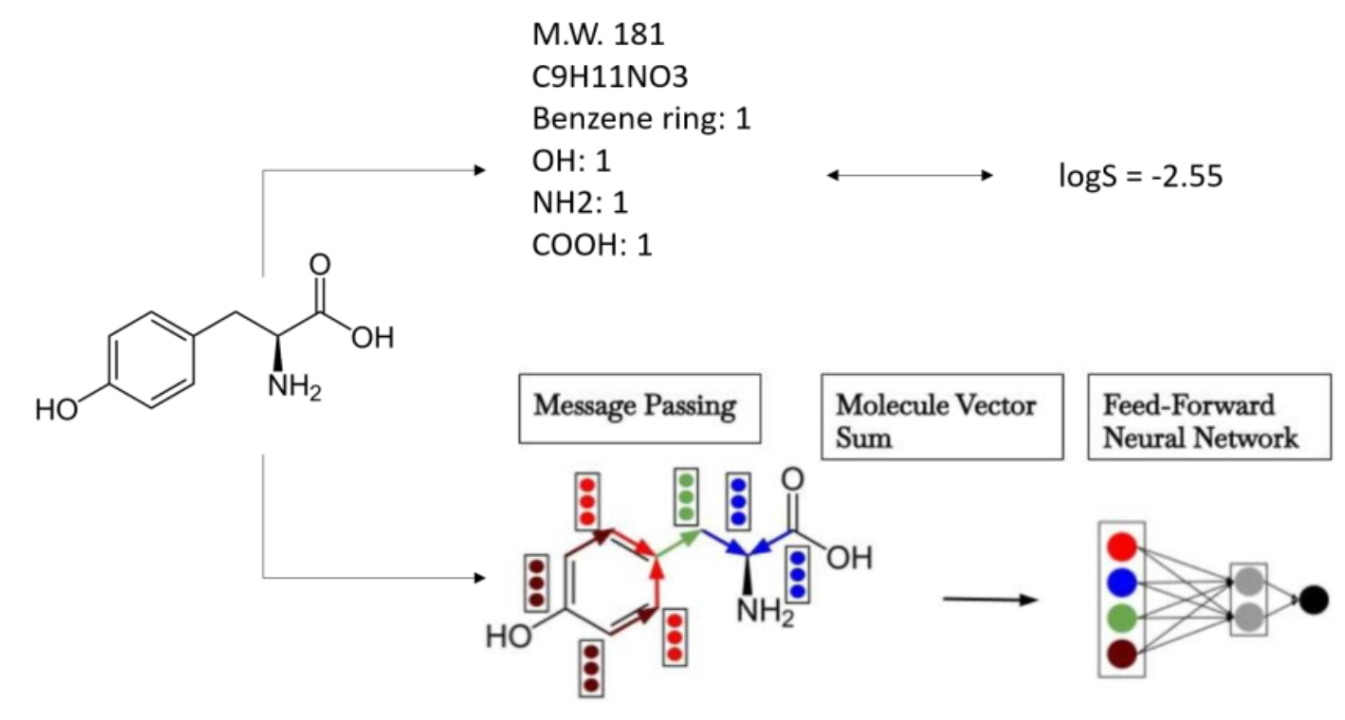}
  \caption{\textbf{Configuration of the linear regression model (above) and GCNN model (below), using the tyrosine molecule as an example.} The linear regression model relies on human-engineered features, including molecular weight (MW) and the count of functional groups, to predict experimental solubility (logS), whereas the GCNN utilizes features acquired via message across a graph.}
  \label{fig:fig1}
\end{figure}

\section{Methods}
\label{sec:methods}
Two ML models were applied in this study: A linear regression model and a GCNN model (Figure 1). 

In the linear regression model, we incorporated the molecular weight, total atom counts, and functional group counts as features to establish a multivariable regression with the experimental solubility values (logS). The features were directly obtained from the molecular structure using the RDKit module \cite{landrum2016rdkit} in SMARTS notation \cite{daylight2007smarts}. We also used L1 regularization with an alpha value of 0.01.

As for the GCNN, we employed the Chemprop model \cite{heid2023chemprop}, which converts the atoms and bonds in the molecules into one-hot encoding, subsequently concatenating them into one tensor representing each individual atom or bond. Chemprop could construct three distinct tensors: one that maps each atom to its corresponding bonds (a2b), another that maps each bond to its corresponding atom (b2a), and a third that maps each bond to its reverse bond (b2revb). Subsequently, it combines each atom tensor into a unified vector and each bond tensor into another consolidated vector. Employing these five tensors, Chemprop identifies the neighboring bonds for each bond and aggregates their vector representations. Finally, the model appends this sum to the vector representations of both the bonds and atoms. These summated vector representations of individual bonds are then combined to generate one feature vector for the entire molecule, which enters a standard feed-forward neural network with a single output (logS).

We tested both models on three different datasets: the Delaney, Huuskonen, and AqSolDB. To evaluate overall accuracy, we employed 5-fold cross-validation within each dataset, utilizing the root mean square error (RMSE) of the parity plots to assess the overall accuracy of the predictions.

\section{Results}
\label{sec:results}
\subsection{Predicting Solubility}

The parity plots for each model on different datasets are plotted in Figure 2, and the root-mean-square deviations (RMSE) are listed in Table 1. 

\begin{table}[!htb]
 \caption{Performance of the Linear Regression Model and GCNN Model on Three Solubility Datasets}
  \centering
  \begin{tabular}{lccc}
    \toprule
    Dataset & Size & RMSE, Linear Regression Model & RMSE, GCNN Model \\
    \midrule
    Delaney & 1127  &  1.13  &  0.59\\
    Huuskonen & 1282  &  1.08  &  0.49\\
    AqSolDB & 9982  &  1.83 & 0.76\\
    \bottomrule
  \end{tabular}
  \label{tab:table1}
\end{table}

\begin{figure}[!htb]
  \centering
  \includegraphics[scale = 1.0]{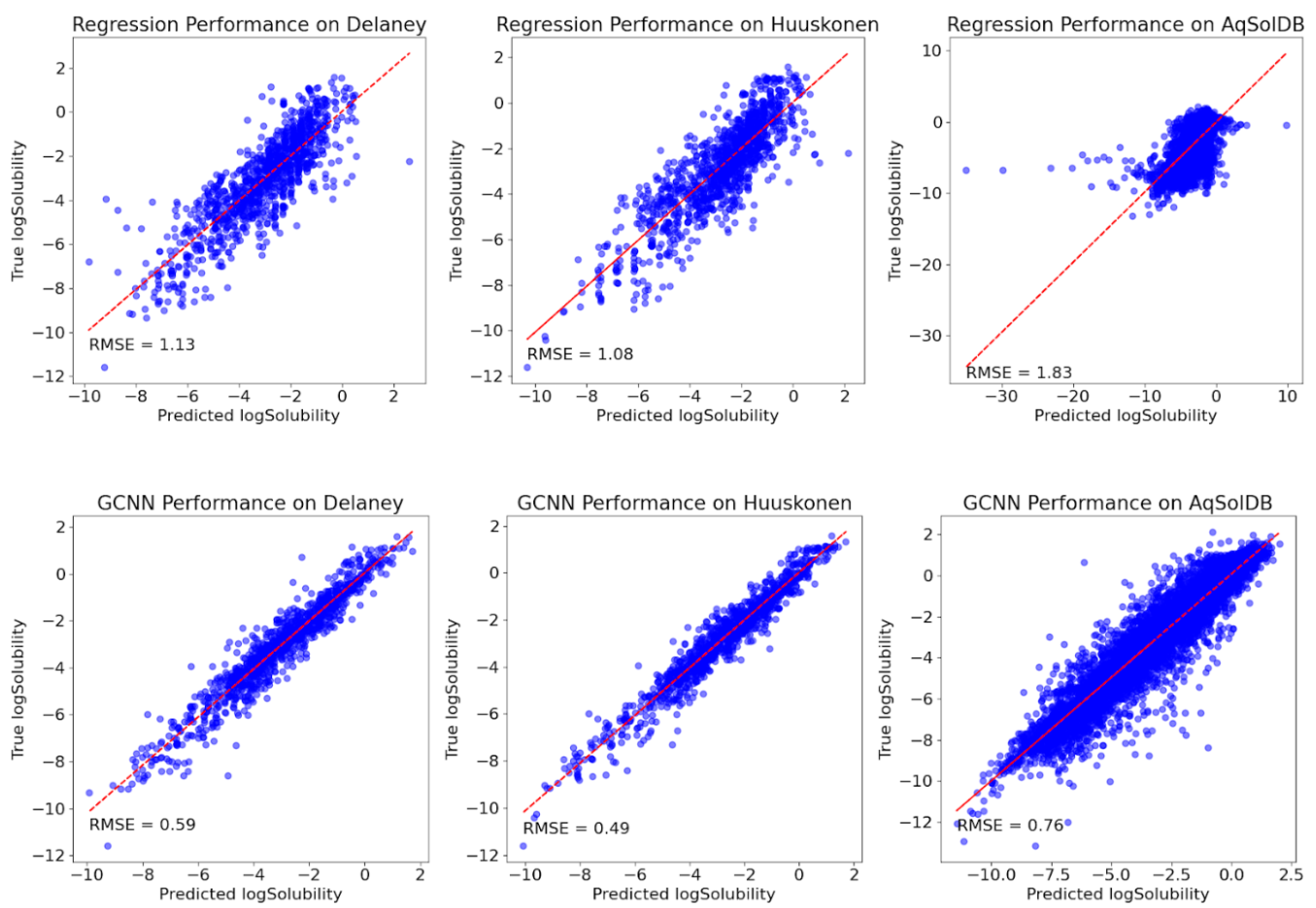}
  \caption{\textbf{Parity plots for the Delaney, Huuskonen, and AqSolDB datasets using linear regression model (top) and GCNN model (bottom).} Predictions are shown from the validation folds of 5-fold cross validation. Lines of best fit are shown in red.
}
  \label{fig:fig2}
\end{figure}

Across all three datasets, both the linear regression model and the GCNN model produced reasonably accurate predictions, with the majority of predicted values falling within 1 log unit of the actual values, consistent with findings in similar studies \cite{boobier2020machine, ye2021prediction}. Notably, both models exhibited their best performance on the Huuskonen dataset and the least optimal performance on the AqSolDB dataset, characterized by the lowest RMSEs. By analyzing the outliers in Figure 2C, those data points correspond to ionic compounds such as compounds containing Zr\textsuperscript{4+}, Al\textsuperscript{3+}, and Zn\textsuperscript{2+}.

The GCNN model, on the whole, outperformed the linear regression model, particularly when dealing with larger and diverse datasets, a result that aligns with the complexity and effectiveness of CNNs observed in various fields, including computer vision. However, this doesn't diminish the value of linear regression. Considering errors stemming from experimental conditions like pH and temperature, both models exhibit sufficient capabilities for drug design purposes.

\subsection{Understanding the relationship between molecular structure and solubility}

In contrast to the GCNN approach, which operates as a "black box", the linear regression model provides a relatively transparent depiction of the direct relationship between the input features and the solubility property of interest. Through feature importance analysis, we can visualize how each feature influences the final results. The significance of different atom species is presented in Figures 3.

\begin{figure}[!htb]
  \centering
  \includegraphics[scale = 1.1]{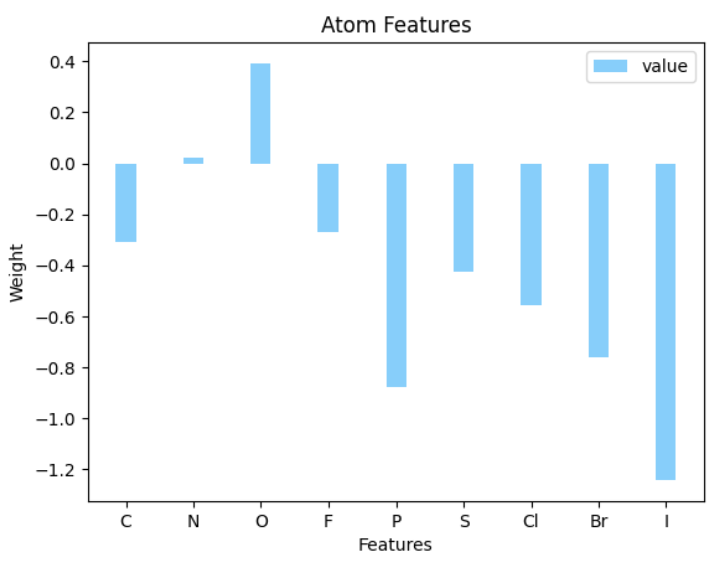}
  \caption{\textbf{The linear regression weights of each type of atom feature for the Delaney dataset.} Positive weights indicate features contributing to a relative increase in solubility, whereas negative weights indicate features which contribute to a relative decrease in solubility.
}
  \label{fig:fig3}
\end{figure}

Solubility hinges on the intermolecular forces between solute and solvent (water) molecules. In essence, polar molecules with more hydrogen bonds, whether as donors or acceptors, tend to exhibit higher solubility in aqueous solutions. The feature analysis results presented here provide a quantitative perspective on these conclusions. For instance, oxygen (O) atoms exert a strong positive influence on solubility because they not only increase the overall polarity of the organic molecules but also have the capacity to form hydrogen bonds with solvent water molecules. Conversely, halogens have a negative impact on solubility which can be quite counter intuitive \cite{huibers1998correlation}. It is generally believed that halogen atoms, especially F and Cl can be hydrogen bond acceptors. However in reality, halogen atoms attaching to carbon chains could not form hydrogen bonds with water molecules. This underscores the pivotal role of hydrogen bonds in aqueous solubility, often surpassing the limited polarity they enhanced. It is also interesting to observe the trend of negative impact on solubility that I > Br > Cl > F, where heavier molecules are less likely to be soluble, while halogenated molecules are more likely soluble in hydrophobic solvents \cite{alsaleem2015solubility}. This trait carries profound implications for drug delivery across cell membranes, making this extended exploration of solubility an area of considerable importance within the pharmaceutical field for further investigation.

\begin{table}[!htb]
 \caption{Performance of the Linear Regression Model with Only Atom Feature and with Atom and Functional Group Features on Three Solubility Datasets}
  \centering
  \begin{tabular}{lccc}
    \toprule
    Dataset & Size & RMSE, Atom Features Only & RMSE, Atom and Functional Group
Features \\
    \midrule
    Delaney & 1127  &  1.13  &  0.96\\
    Huuskonen & 1282  &  1.08  &  0.89\\
    AqSolDB & 9982  &  1.83 & 1.73\\
    \bottomrule
  \end{tabular}
  \label{tab:table2}
\end{table}

Notably, the inclusion of functional group counts on top of atom counts yields a substantial improvement in the RMSE, as demonstrated in Table 2. This implies that the same type of atom when integrated into different functional groups, can exert varying effects on solubility. For instance, certain atoms like N, S, and P have the capacity to form diverse functional groups, which in turn may have either positive or negative impacts on solubility. The impact of different functional groups to aqueous solubility are shown in Figure S1.

It is essential to recognize that the machine learning models in this study can only predict a single solubility value for a given molecular structure. In reality, scientists contend with high-dimensional data encompassing a range of solubility values under varying conditions for each compound, as well as other physical and chemical properties. Tackling this complexity necessitates extensive data collection, cleaning, and algorithm development efforts. Ultimately, we anticipate that a sophisticated neural network-based model, coupled with interpretable feature analysis, will emerge as the preferred tool of choice, surpassing the simple linear regression approach.

\subsection{From solubility prediction to drug design}

As discussed above, simple solubility models have proven effective for high-throughput screening, even with the long-established GSE. Yet, the broader significance of solubility studies emerges in their capacity to inform and influence future drug design. This presents a reverse perspective: When endeavoring to create a drug molecule with specific solubility values or other desired physical attributes, the pivotal question becomes, which functional groups should be incorporated?

Using the insights gained from feature importance analysis in this study, it is possible to develop a general understanding of which functional groups to incorporate. For instance, to enhance aqueous solubility, introducing an OH group to a side chain can be an effective strategy. At the same time, for improving the ability to permeate cell membranes, the inclusion of a halogen atom might be the most suitable choice. However, in real-world scenarios where multiple factors must be considered simultaneously, the complexity of human decision-making can be quickly overwhelmed. This is precisely where the GCNN model proves invaluable. By leveraging a well-trained neural network that establishes connections between defined molecular substructures and their associated properties, the coupling of the GCNN with a molecular generative model \cite{merz2020generative} has the potential to enable the generation of viable drug candidates with desired properties on a larger scale. This approach will likely drive the next generation of high-throughput screening in the pharmaceutical industry.

\section{Conclusion}
\label{sec:discussion}
In this investigation, we tried to predict the drug molecules aqueous solubility by applying two distinct models: a linear regression model with human-engineered features, and a GCNN model. Both models exhibit commendable predictive accuracy across diverse datasets, with the GCNN delivering superior overall performance. Nonetheless, the linear regression model offers a valuable lens into the intricate interplay between specific features and solubility, shedding light on the significance of certain atoms, functional groups, and hydrogen bonds in the process. The integration of a GCNN model with feature analysis represents a promising avenue for future research in this domain.

\section{Acknowledgements}
\label{sec:acknowledgements}
We would like to thank Kyle Swanson for his valuable suggestions and feedback for this study.

\clearpage

\section{Supplemental Figures}
\begin{figure}[!htb]
  \centering
  \includegraphics[scale = 0.8]{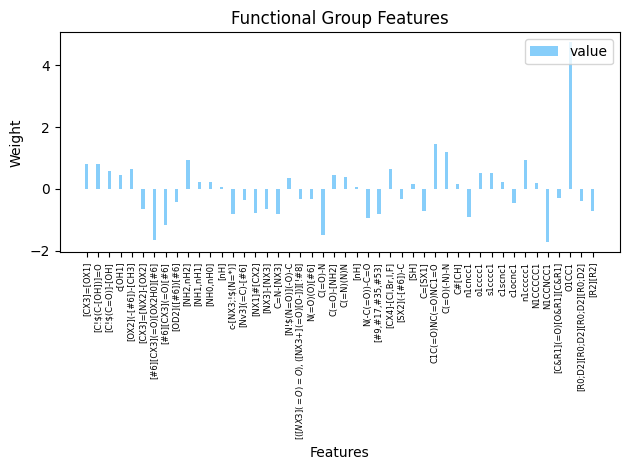}
  \caption{\textbf{The weights of each type of functional group feature in SMARTS notation.} Positive weights indicate features contributing to a relative increase in solubility, whereas negative weights indicate features which contribute to a relative decrease in solubility.
}
  \label{fig:fig4}
\end{figure}

\clearpage
\bibliographystyle{unsrt}  
\bibliography{references}

\end{document}